\documentclass{pasj00}
\onecolumn
\begin{document}
\SetRunningHead{J.\ Fukue}
{Observational Appearance of Black Hole Winds}
\Received{2008/00/00}
\Accepted{2008/00/00}

\title{Observational Appearance of Black Hole Winds}

\author{ Jun \textsc{Fukue} and
Naoko \textsc{Sumitomo}} 
\affil{Astronomical Institute, Osaka Kyoiku University, 
Asahigaoka, Kashiwara, Osaka 582-8582}
\email{fukue@cc.osaka-kyoiku.ac.jp}



\maketitle


\begin{abstract} 
We examine the observational appearance of
an optically thick, spherically symmetric, relativistic wind
(a black hole wind).
In a relativistic flow,
the apparent optical depth becomes small 
in the downstream direction,
while it is large in the upstream direction
due to the relativistic effect.
Hence, the apparent photosphere of the relativistic flow
depends on the flow velocity and direction as well as the density distribution.
We calculate the temperature distribution of the apparent photosphere
of the optically thick black hole wind,
where the wind speed is assumed to be constant and
radiation dominates matter,
for various values of the wind speed and mass-outflow rate.
We found that the limb-darkening effect is strongly enhanced
in the relativistic regime.
We also found that the observed luminosities of the black hole wind
become large as the wind speed increases,
but do not depend on the mass-outflow rate.
\end{abstract}


\section{Introduction}
The {\it observational appearance} is essential
in astrophysical problems,
since we obtain various imformation from only observation
of celestial bodies except for inner solar system objects.
Hence, we usually calculate
emission spectra, light curves, and imaging properties,
after we construct a model of astrophysical phenomena.

In recent years
relativistic astrophysical phenomena become more and more important,
relating to quasars and active galaxies,
micro-quasars and black-hole binaries 
(Mirabel \& Rodr\'\i guez 1999; Fender et al. 2004; Fabrika 2004),
gamma-ray bursts (M\'esz\'aros 2002; Daigne 2002; Piran 2008), and so on.
In these objects
the central engine is believed to be
a black hole surrounded by an accretion disk (see, e.g., Kato et al. 2008).
The gravitational energy released in the mass accretion process
is converted into the tremendous radiation, magnetic, and kinetic energies.
As a result,
the relativistic jets and winds often blow off from
these systems --- a {\it black hole wind}.

Up to now,
the dynamics of relativistic jets and winds
have been extensively investigated (e.g., Beskin et al. 2004).
The flows would be accelerated to relativistic speed
via hydrodynamical, magnetohydrodynamical,
and radiation hydrodynamical forces.
For example, spherically symmetric, relativisitic winds 
under radiatively driven mechanisms
have been investigated by several researchers
(e.g., Castor 1972; Ruggles \& Bath 1979; Mihalas 1980; 
Quinn \& Paczy\'nski 1985; Turolla et al. 1986; Paczy\'nski 1986, 1990; 
Akizuki \& Fukue 2008, 2009),
and their models have applied to neutron star winds and gamma-ray bursts.

On the other hand, the observational appearance of such relativistic flows
has never been considered except for a few cases (Sumitomo et al. 2008).
Of course, it is well known that
the relativistic flows suffer from the Doppler boosting;
e.g.,
the observed frequency-integrated intensity
is enhanced by the fourth power of the Doppler factor.
Hence, in the global (rough) view point,
the total observed luminosity
is usually corrected under such a relativistic effect.
However, in the local (detailed) view point,
the observational appearance of relativistic flows
has not been considered yet.

Observational appearence of relativistic flows
would be affected by various relativistic effects.
In particular,
we carefully treat an {\it apparent photosphere}
of optically thick relativistic flows
under the viewpoint of the relativistic radiation transfer.
First, the optical depth is affected by the mass-outflow rate.
That is,
if the mass-outflow rate of winds exceeds the critical rate,
the optical depth of the wind becomes larger than unity
(see, e.g., King \& Pounds 2003).
As a result,
such a massive wind may form a ``photosphere'',
and the location of the photosphere
 depends on the wind velocity and mass-outflow rate.
Hence, in order to calculate spectra and luminosities of winds,
we should determine the location of the photosphere 
and the temperature distribution there.

Furthermore,
the apparent optical depth is modified in the relativistic flow.
Namely,
Abramowicz et al. (1991) found that, in the relativistically moving media,
the apparent optical depth becomes small 
in the downstream direction,
while it is large in the upstream direction
due to the relativistic effect.
As a result,
the apparent photosphere is futher modified;
the shape of the photosphere appears convex in the non-relativistic case, 
but concave for relativistic regimes. 
Hence, in order to calculate spectra and luminosities 
of {\it relativistic} winds,
we should determine the location of the photosphere 
and the temperature distribution there
under the viewpoint of the {\it relativistic radiation transfer}.

Recently,
Sumitomo et al. (2008) has firstly examined
the observational appearance of 
spherically symmetric, relativistic massive winds.
They assumed a simple optically thick wind,
which has a constant speed, 
mainly consists of baryon, and adiabatically expands.
They found that
the wind luminosities increase with the wind speed
due to the relativistic boost,
but decrease with the mass-outflow rate
due to the temperature decrease in the outer region.

Since Sumitomo et al. (2008) determined the temperature of the flow
under the assumption of the adiabatic expansion of the baryon dominated fluid,
the resultant radiative luminosity is not constant spatially,
but decreases (cf. Paczy\'nski 1990).
In the radiation dominated fluid, on the other hand,
the temperature and luminosity distribution may change;
e.g., the luminosity is constant (cf. Paczy\'nski 1986).
In this paper
we thus consider the observational appearance of a black hole wind
under a continuous, steady flow picture of the radiation dominated fluid,
where the luminosity is spacially constant.

%

In section 2 the present wind model and the calculation method
are briefly described.
In section 3 the results are shown.
Final section is devoted to concluding remarks.
In the appendix the opposite problem,
the observational appearance of a black hole accretion,
is briefly examined.


\section{Wind Model and Calculation Method}

In this section
we describe the optical depth,
the present wind model,
and the calculation method.

\subsection{Optical Depth and Apparent Photosphere}

We assume that a spherically symmetric, relativistic wind blows off from a central object. 
As a central object,
we assume a non-rotating black hole (Schwarzschild black hole),
 and the Schwarzschild radius is defined by $r_{\rm g} = 2GM/c^2$,
 where $G$, $M$, and $c$ represent the gravitational constant, 
 a black hole mass, and the speed of light, respectively.
We use the spherical coordinates $(R, \theta, \varphi)$ and
the cylindrical coordinate $(r, \varphi, z)$,
whose $z$-axis is along the line-of-sight  (see figure 1). 

We define an ``apparent photosphere'' of the wind as the surface, where
 the optical depth $\tau$ measured from 
an observer at infinity becomes unity.  
Schematic picture  of the present calculation is presented in figure 1.

Lets us consider a small distance $ds$ along the light path. 
Due to the relativistic effect (Lorentz-Fitzgerald contraction),
the mean free path of photons in the fixed frame, $\ell$,
is related to that in the comoving frame, $\ell_0$, by
\begin{equation}
    \ell = \frac{1}{\gamma (1- \beta \cos \theta)} \ell_0,
\end{equation}
where $\beta$ ($=v/c$) is the wind speed $v$
normalized by the speed of light,
$\gamma$ ($=1/\sqrt{1-\beta^2}$) the Lorentz factor,
and $\theta$ the viewing angle measured from the $z$-axis.
Here and hereafter, quantities with subscript ``0'' 
refer to physical quantities measured in the comoving frame. 
Then, the optical depth in the fixed frame is given by 
\begin{equation}
  d\tau = d\tau_0 = \kappa_0 \rho_0 ds_0
        = \gamma (1- \beta \cos \theta) \kappa_0 \rho _{\rm 0} ds,
  \label{eq:dtau}
\end{equation}
where the opacity $\kappa_0$ is assumed to be electron scattering
(Abramowicz et al. 1991). 
Thus, the optical depth strongly depends on the viewing angle $\theta$ 
as well as the flow speed $v$.
It is obvious that the optical depth is smallest 
in the downstream direction at $\theta=0$, 
where the photons move in the same direction with the fluid,
while it becomes largest in the upstream direction at $\theta=\pi$,
where the photons move in the opposite direction to the fluid. 

From equation (\ref{eq:dtau}), the integrated optical depth $\tau_{\rm ph}$ from an observer at infinity is calculated as 
\begin{equation}
\tau_{\rm ph} = \int_{z_{\rm ph}}^\infty 
\gamma (1- \beta \cos \theta) \kappa_0 \rho _{\rm 0} ds=1,
\label{eq:tauph}
\end{equation}
where ${z_{\rm ph}}$ is the height 
of the apparent photosphere from the equatorial plane.
Abramowicz et al. (1991) showed that 
the photosphere of a highly relativistic wind 
is much closer to the source, roughly by a factor $\gamma^2$. 
Although the non-relativistic and moderately relativistic winds 
have convex photospheres, the photospheres of relativistic winds 
becomes concave for $\beta > 2/3$ (see figure 2 later).

\begin{figure}
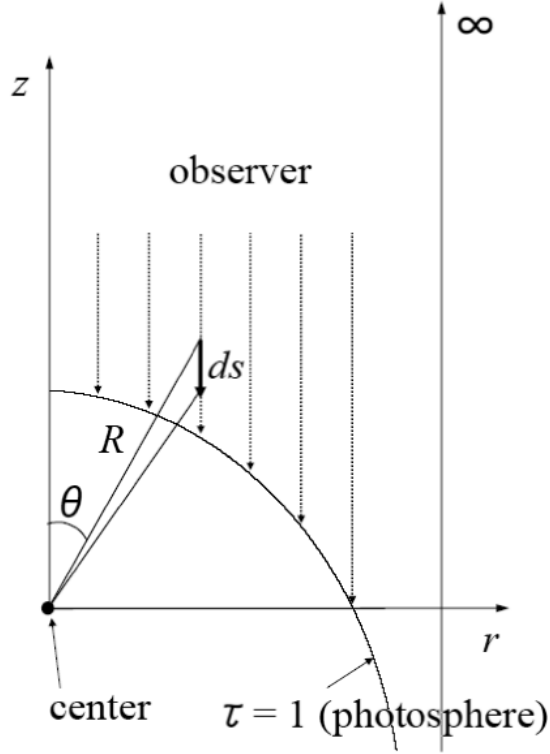

  \begin{center}
  \FigureFile(80mm,80mm){figure01.eps}
@\end{center}
\caption{Schematic picture of the present calculation.
The spherical wind is assumed to blow off from the origin
at a relativistic speed.
The observer is located at infinity in the $z$-direction.
An apparent ``photosphere'' is defined so that
the optical depth measured from the infinity becomes unity.
}
\end{figure}

\subsection{Density Distribution}

From continuity equation, for the spherically symmetric stationary wind,
the rest mass density $\rho_{\rm 0}$ measured in the comoving frame
varies as  
\begin{equation}
  \rho_{\rm 0} = \frac{\dot{M}}{4 \pi \gamma v} \frac{1}{R^{2}},
\end{equation}
 where 
$\dot{M}$ is the constant mass-outflow rate and 
$R=\sqrt{r^2 +z^2}$ is a distance from the central object, 

We suppose that the black hole wind is well accelerated
at the very center of the flow,
where the optical depth is sufficiently large,
to reach a constant terminal speed in the observational region
(cf. Paczy\'nski 1990).
Hence, in this paper the wind velocity $v$ is assumed to be constant,
and the density varies as $\rho_0 \propto 1/R^2$. 

\subsection{Temperature Distribution and Luminosity}

In order to determine the temperature distribution of the wind,
Sumitomo et al. (2007) assumed the adiabatic expansion of the baryonic gas.
In the present model, on the other hand,
we consider the radiation dominated fluid, and
we assume that the spherical wind continuously blows off
at a constant terminal speed,
there is no heating and cooling in such a region,
and therefore, the total energy flux is conserved;
the relativistic Bernoulli equation holds.
The relativistic Bernoulli equation in the spherical geometry is
\begin{equation}
\dot{M}c^2 \frac{\varepsilon + p}{\rho c^2}\gamma \sqrt{g_{00}}
+ 4\pi R^2 g_{00} F = {\rm constant},
\label{Ber1}
\end{equation}
where $\varepsilon$ is the gas internal energy per unit proper volume,
$p$ the gas pressure,
$g_{00}$ ($=1-r_{\rm g}/R$) the metric,
and $F$ the radiative flux in the fixed frame.
The radiative flux $F$ in the fixed frame is expressed as
\begin{equation}
   F = \gamma^2
   \left[ \left( 1+\beta^2 \right) F_0
   + \beta \left( cE_0 + cP_0 \right) \right],
\end{equation}
using the radiation energy density $E_0$,
the radiative flux $F_0$, and 
the radiation pressure $P_0$ in the comoving frame.
This equation (\ref{Ber1}) is rather general form
(cf. Paczy\'nski 1990; Kato et al. 2008).

On the contrary to Sumitomo et al. (2007) and Paczy\'nski (1990),
we consider the radiation dominated fluid (cf. Paczy\'nski 1986).
Hence, we ignore the baryonic part of equation (\ref{Ber1}) 
and have
\begin{eqnarray}
   L &=& 4\pi R^2 g_{00} F
\nonumber \\
     &=& 4\pi R^2 g_{00} \gamma^2
   \left[ \left( 1+\beta^2 \right) F_0
   + \beta \left( cE_0 + cP_0 \right) \right]
\nonumber \\
   &=& {\rm constant},
\label{Ber2}
\end{eqnarray}
where 
\begin{eqnarray}
   L_{\rm dif}
     &=& 4\pi R^2 g_{00} \gamma^2
      \left( 1+\beta^2 \right) F_0,
\\
   L_{\rm adv}
     &=& 4\pi R^2 g_{00} \gamma^2
    \beta \left( cE_0 + cP_0 \right),
\end{eqnarray}
are the diffusive and advective luminosities, respectively.
It should be noted that
this equation (\ref{Ber2}) is equivalent with,
e.g., equation (3b) of Paczy\'nski (1990),
except for the ignorance of the baryonic part.

We further suppose that
the gas locally emits blackbody radiation in the comoving frame
at a temperture $T_0$:
$F_0=\sigma T_0^4$ and $E_0=3P_0=aT_0^4$,
where
$\sigma$ is the Stephan-Boltzmann constant and
$a$ ($=4\sigma/c$) is the radiation constant.
Hence, 
the luminosity in the fixed (inertial) frame is expressed as
\begin{equation}
   L = 4\pi R^2 F
   = 4\pi R^2 \gamma^2 \left( 1+\frac{16}{3}\beta + \beta^2 \right)
   \sigma T_0^4 = \dot{E}',
\end{equation}
where we set $g_{00}=1$ at large $R$.
It should be noted that 
this equation is a cousin of equation (3) in Paczy\'nski (1986)
for a fireball model of gamma-ray bursts,
where the diffusion luminosity is ignored and $\beta \sim 1$.
Since the flow speed is assumed to be constant,
the luminosity in the comoving frame becomes constant:
\begin{equation}
   L_0 = 4\pi R^2 \sigma T_0^4 = \dot{E},
\end{equation}
or the temperature $T_0$ of the wind gas in the comoving frame
varies as
\begin{equation}
   T_0 = \left( \frac{\dot{E}}{4\pi \sigma R^2} \right)^{1/4},
\end{equation}
where
$\dot{E}$ is the energy-outflow rate (luminosity) in the comoving frame.
In the Paczy\'nski's (1986) fireball model, 
where the self-similar solution is adopted and
the Lorentz factor is proportional to $R$,
the temperature is proportional to $1/R$. 
The validity of this assumption will be discussed 
in the last section.

Finally, the observed temperature $T_{\rm obs}$ 
measured by an observer at infinity is expressed by
\begin{equation}
T_{\rm obs} = \frac{1}{1+z} T_0 = \frac{1}{\gamma (1- \beta \cos \theta)}T_{\rm 0},
\end{equation} 
where $z$ is the redshift via longitudinal and transverse Doppler effects.
Using this observed temperature,
we can obtain the observed luminosity at infinity by 
\begin{equation}
L = \int_0 ^{r_{\rm out}} \sigma T_{\rm obs}^4 2\pi rdr,
\end{equation} 
where $r_{\rm out}$ takes a sufficiently large radius.


\section{Results}

We determined the apparent photosphere for various $\beta$ 
via equation (\ref{eq:tauph}),  and obtained the temperature and luminosity 
on the surface of the photosphere in the comoving and inertial frames. 
In the present calculation,
we bear in mind the case for microquasars,
although the present results are also important 
for active galactic nuclei and gamma-ray bursts.
Hence,
the black hole mass is fixed as $M=10 M_{\odot}$. 
The input parameters are then the wind speed $\beta$, 
the normalized luminosity $\dot{e}$
($\equiv \dot{E}/L_{\rm E}$),
where $L_{\rm E}$ is the Eddington luminosity of the central object, and
the normalized mass-outflow rate $\dot{m}$ ($=\dot{Mc^2}/L_{\rm E}$).
Of these, we set $\dot{e}=1$ for simplicity.

\subsection{Location of the Apparent Photosphere}

Figure 2 shows the location of the apparent photosphere seen 
by the observer at infinity in the $z$-direction for various wind velocities.  
In the low speed regime,
the photosphere near the $z$ axis is close to the center,
 while the photosphere far away from the  $z$ axis is far from the center. 
This is the usual limb-darkening effect of a spherically expanding wind.

In the high speed regime, on the other hand,
the shape of the apparent photophere changes,
because the optical depth depends on the angle $\theta$ and 
the wind velocity $v$.
In particular, when the wind blows off at highly relativistic speed 
($\beta \geq 2/3$),
 the photosphere looks like a concave shape.

Our results consistent with the analytical results 
by Abramowicz et al. (1991), 
 but we note that the units of our coordinate 
is the Schwarzschild radius $r_{\rm g}$. 

It should be noted that from equations (3) and (4)
the height $z_{\rm ph}(0)$ of the apparent photosphere
at $r=0$ ($\theta = 0$) becomes
\begin{equation}
   \frac{z_{\rm ph}(0)}{r_{\rm g}}
   = \frac{1-\beta}{\beta} \dot{m}.
\end{equation}
Hence, the size of the apparent photosphere of a black hole wind
roughly expands in proportion to the mass-outflow rate.
In addition, for the extremely relativistic case of $\beta \sim 1$,
the size is proportional to $1/\gamma^2$,
as Abramowicz et al. (1991) showed.

\begin{figure}
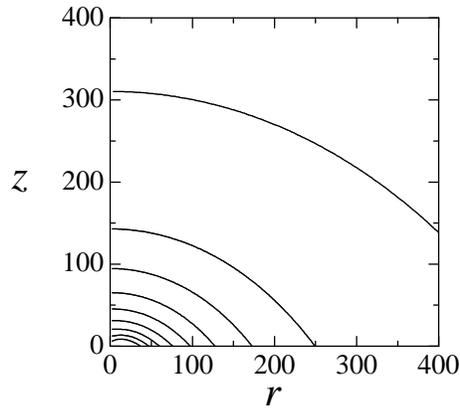

  \begin{center}
  \FigureFile(80mm,80mm){figure02.eps}
  \end{center}
\caption
{
Location of the apparent photosphere for various wind velocity $\beta$
in the case of $\dot{m}=100$.
The wind velocity is varied from 0.1 (outer region) to 0.9 (central region) in steps of 0.1. 
The units of the $r$- and $z$-axis is the Schwarzschild radius $r_{\rm g}$.
}
\end{figure}

\subsection{Apparent Temperature}

In figures 3 and 4 we show the temperature distributions 
in the comoving and inertial frames, respectively. 
In general,
the wind photosphere looks brightest at the central part, 
and the surroundings are gradually dim as increasing radius.
This is due to the limb darkening effect seen in the usual spherical wind.

In the relativistic wind considered here, however,
this limb-darkening effect is remarkably enhanced.
This is
due in part to the relativistic Doppler and aberration effects,
and due in part to the fact that
the observed photosphere shrinks as the velocity increases and
we can see deep inside the wind.

In the comoving temperature (dashed curves in figure 3),
the temperature becomes heigher and heigher
as the flow speed increases.
This is just the observed photosphere shrinks as the velocity increases,
and this relativistic effect is significant in the central part.

Furthermore,
comparing the temperature distributions in the comoving and inertial frames,
we see that
the central temperature in the inertial frame is 
much higher than that in the comoving frame,
and the outer temperature much lower. 
This is just the relativistic Doppler and aberration effects.
That is, the observed temperature increases as $\theta$ approaches zero
 because of the longitudal Doppler effect (highly beamed emission).
This effect also becomes remarkable as the velocity increases. 

In addition,
the temperature at the apparent photosphere decreases
as the mass-outflow rate increases (figure 3),
since the apparent photosphere becomes small as the mass-outflow rate increases
and the energy-loss rate is set to be the Eddington rate.

\begin{figure}
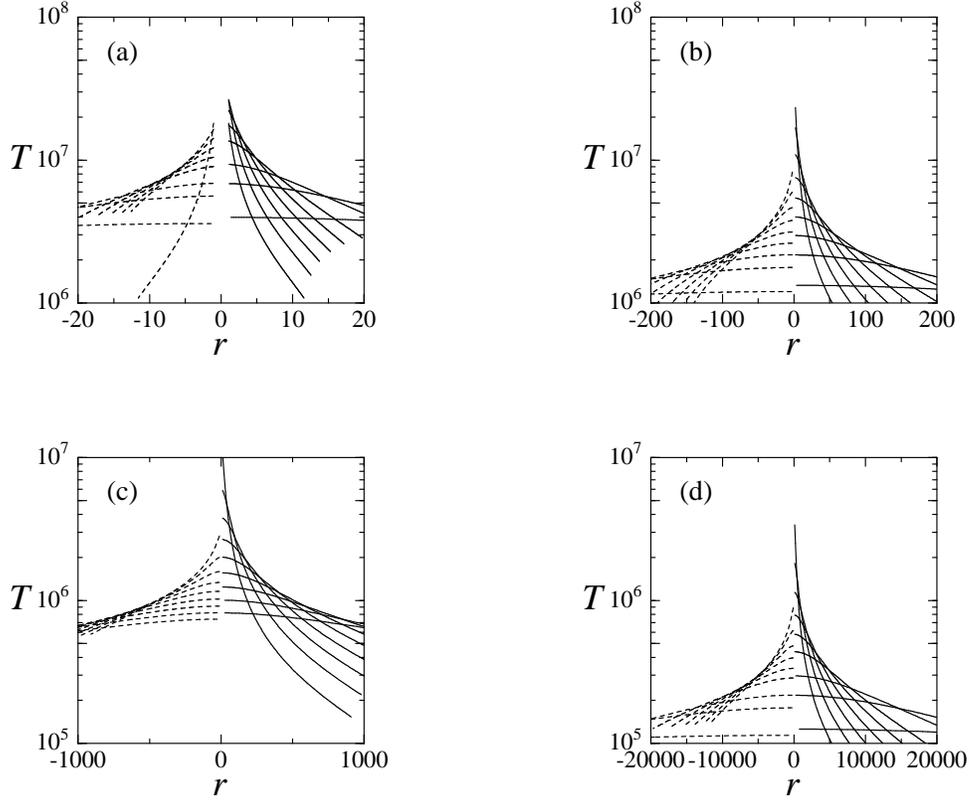

  \begin{center}
  \FigureFile(75mm,75mm){figure03a.eps}
  \FigureFile(75mm,75mm){figure03b.eps}
  \FigureFile(75mm,75mm){figure03c.eps}
  \FigureFile(75mm,75mm){figure03d.eps}
\end{center} 
\caption{
Temperature distribution at the apparent photosphere
as a function of radius $r$.
The dashed curves in the leftpart are the temperatures in the comoving frame,
while the solid ones in the rightpart are those 
in the fixed (observer's) frame.
The mass-outflow rates $\dot{m}$ are
(a) 10, (b) 100, (c) 1000, and (d) 10000.
In each panel
the wind velocity $\beta$ is
0.1 to 0.9 in steps of 0.1 from flat to steep curves.
}
\end{figure}

\begin{figure}
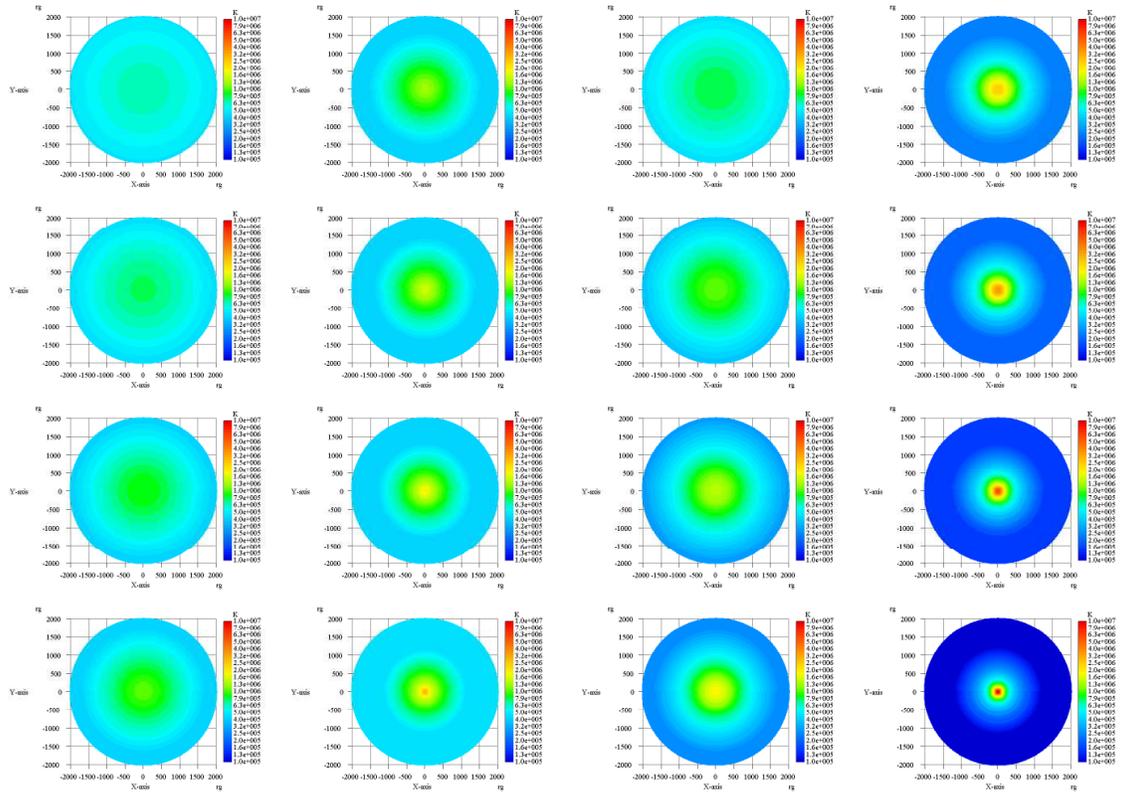

  \begin{center}
  \FigureFile(75mm,75mm){figure04a.eps}
  \FigureFile(75mm,75mm){figure04b.eps}
  \end{center} 
\caption{
Temperature distribution at the apparent photosphere
viewed by an observer at infinity in the $z$-direction.
The left panel shows the temperature in the comoving frame,
while the right panel is that in the fixed (observer's) frame.
In each panel
the wind velocity $\beta$ is 0.2 to 0.9 in steps of 0.1 from top-left
to bottom-right. 
The mass-outflow rate $\dot{m}$ is 1000.
}
\end{figure}

\subsection{Apparent Luminosity}

In figure 5
we show the luminosities of the relativistic winds
in the present model
as a function of the wind speed $\beta$
for several mass-outflow rates $\dot{m}$.
Solid curves represent the observed luminosity in the inertial frame,
 and the dashed ones show the comoving luminosity.

As is seen in figure 5,
the comoving luminosity slowly increases as the velocity increases,
although the values are close to the Eddington luminosity.
Furthermore,
the luminosity does not depend on the mass-outflow rate.
These behaviors are quite different from those in Sumitomo et al. (2008),
where, e.g., the luminosity decreases as the mass-outflow rate increases.
The present behavior is natually understood, since
the comoving luminosity is costant without heating and cooling,
and the energy-loss rate is fixed as $\dot{e}=1$
in the present model.
The weak dependence on the velocity is also
the effect of the apparent optical depth in the relativistic regime.

It should be noted that
in the case of $\dot{m}=10$
the observed luminosity drops for large $\beta$.
This is because the apparent optical depth
becomes less than unity in such a regime
of small mass-outflow rate and high velocity.

In addition,
the luminosity observed in the inertial frame is higher than
that in the comoving frame,
and depends on the flow velocity,
but is independent of the mass-outflow rate generally.
This is mainly due to the relativistic boost
of the comoving luminosity under the effect of the apparent optical depth;
i.e.,
the luminosity is observed roughly by a factor of $\gamma^4$
in the inertial frame.
It is emphasized that the comoving luminosity is enhanced about 114 percent  for $\beta=0.1$, but the amplification is about factor 6 for $\beta=0.9$. 
These facts suggest that there is a possibility of overestimation of the observed luminosities for relativistic outflow objects.  

\begin{figure}
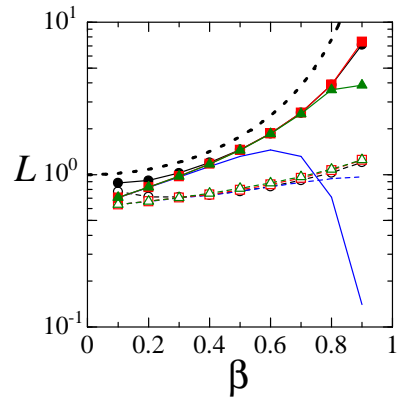

  \begin{center}
  \FigureFile(80mm,80mm){figure05.eps}
  \end{center}
\caption{
Luminosity of a black hole wind as a function of the wind velocity $\beta$
for several mass-outflow rates $\dot{m}$.
Solid curves represent the observed luminosity in the inertial frame,
 and the dashed ones show the comoving luminosity.
The mass-outflow rate ${\dot{m}}$ are 10 (no symbols), 100 (triangles), 
1000 (squares), and 10000 (circles). 
The dotted curve denotes $\gamma^4$.
}
\end{figure}

\section{Concluding Remarks} 

In this paper, we examined the observational appearance of relativistic, 
spherically symmetric black hole winds from the observational point of view.
As for the shape of the apparent photosphere of black hole winds,
we confirmed the results of Abramowicz et al. (1991).
We further calculated the temperature distribution and luminosity 
of the photosphere both in the comoving and inertial frames
under the simple model. 
We found that the limb-darkening effect would strongly modified
in the relativistic regime.
That is, the observed appearance of the photosphere becomes non-uniform
 in spite of  spherical symmetry.
Since the apparent optical depth becomes small with velocity,
we could see deeper inside the wind, as the velocity increases.
This is the {\it relativistic} limb-darkening effect. 
This nature does not depend on the observer's direction.
In addition, the luminosity in the observer's frame is remarkably enhanced by relativistic beaming effects along the observer's direction. 
These two effects mainly work as the luminosity enhancement of the relativistic outflow. 
We suggest that
 if the observed luminosity is used for the evaluation of the black hole mass, 
 then the derived black hole mass will be overestimated. 

\begin{table}
\begin{center}
\caption{Models of Black Hole Winds}
\begin{tabular}{lcccc}
\hline
\multicolumn{1}{c}{Authors} & Velocity & Temperature & Luminosity & Notes \\
\hline
Paczy\'nski (1986)       & $\gamma \propto R$ & $T_0 \propto R^{-1}$ & constant & radiation dominant \\
Paczy\'nski (1990)       & fully solve        & fully solve & $L \sim R^{-2/3}$ & baryon dominant \\
Abramowicz et al. (1991) & constant           & --- & --- & --- \\
Sumitomo et al. (2007)   & constant           & $T_0 \propto R^{-2/3}$ & $L \propto R^{-2/3}$ & baryon dominant \\
Present model            & constant           & $T_0 \propto R^{-1/2}$ & constant & radiation dominant \\
\hline
\multicolumn{4}{@{}l@{}}{\hbox to 0pt{\parbox{85mm}{\footnotesize
}\hss}}
\end{tabular}
\end{center}
\end{table}

As already stated, the purpose of the present paper 
was not to solve the dynamical structure of a black hole wind,
but to examine its observational appearance.
In order to clarify the position of the present work,
we compare the conditions of previous and present works (table 1),
and discuss the validity of the assumptions of the present work.

At first, on the velocity field
we assumed that the velocity is constant.
At the very center of the flow, of course,
the wind is accelerated and the velocity increases.
However, near to the photosphere
the wind generally reaches a constant teminal speed,
e.g., as shown in Paczy\'nski (1990).
Hence, the present assumption of constant velocity is valid,
except that the mass-outflow rate is small,
where the deep inside of the wind can be seen.

On the temperature distribution,
we considered the radiation dominated fluid
and assumed that the total energy flux is conserved.
As a result,
the luminosity in the comoving frame is constant and
the comoving temperature varies as $T_0 \propto R^{-1/2}$.
The assumption of the constant luminosity is same as Paczy\'nski (1986),
although the diffusive luminosity is ignored in Paczy\'nski (1986).
In Sumitomo et al. (2007), on the other hand,
we considered the baryonic fluid
and assumed the adiabatic expansion.
As a result,
the comoving temperature varies as $T_0 \propto R^{-2/3}$
and the luminosity also varies as $L \propto R^{-2/3}$.
In Paczy\'nski (1990), where the baryonic component is important,
the advective luminosity inside the photosphere
dominates the diffusive luminosity and roughly changes
as $L_{\rm adv} \sim R^{-2/3}$.
In the astrophysical situations and applications,
the baryonic winds correspond to
the neutron star and black hole winds
with sufficient amount of normal plasmas.
In luminous black hole wind with less amount of baryon
or in gamma-ray burst fireball with electron-positron plasmas,
on the other hand,
the radiative component would dominate the baryonic one.

Finally, we assumed that
the gas locally emits the blackbody radiation in the comoving frame.
In the deep inside the photosphere
this LTE assumption is valid,
whereas it is violated in the optically thin outer region.
Due to the relativistic effect,
the apparent photosphere locates inside the wind.
Hence, the present LTE assumption would hold,
at least marginally.

In order to compare with observational data,
 we need more strict treatments of a wind model,
 e.g., the effect of general relativity, radiative energy loss in the wind,
 compton processes, acceralation by radiation pressure, and so on. 
We may use a fireball model (e.g., Paczy\'nski 1986),
the numerical model (e.g., Paczy\'nski 1990), or
a radiatively driven relativistic model (e.g., Akizuki \& Fukue 2008).
However, the aim of this paper is to show the possibility of the formation of the photosphere. 
Here, we explicitly show the formation of the photosphere
 in an optically thick wind using a simple spherical wind model.

We should also consider observational spectra
of a black hole wind under the relativistic radiative transfer effect,
including non-blackbody cases.
Strictly speaking, the observed temperature should be evaluated from the temperature on the surface, where the effective optical depth equals to unity,
$\tau_{\rm eff}=\sqrt{\tau_{\rm ff} \tau_{\rm tot}}=\sqrt{\tau_{\rm ff} (\tau_{\rm ff}+\tau_{\rm es})}=1$,
$\tau_{\rm ff}$ and $\tau_{\rm es}$ being the free-free and electron scattering opacities, respectively. 
In a high temperature plasma,
the effective optical depth is often smaller
that the total optical depth,
and the gas becomes scattering dominant.
In such a scattering dominated plasma,
the emergent spectrum is not a simple blackbody
but a modified blackbody..
In addition,
we have used the Thomson cross section for electron scattering.
When the center-of-mass energy of scattering becomes relativistic
($h\nu \sim 100$~keV),
we should use the Klein-Nishina cross section,
which reduces the effective cross section.
In such a highly relativistic regime,
a wind will be much more transparent.
These effects are also left as future problems.


\vspace{10mm}

This work has been supported in part by the Grant-in-Aid 
for Scientific Research of the
Ministry of Education, Culture, Sports, Science, and Technology
(18540240 JF).

\appendix

\section*{Observational Appearance of Black Hole Accretion}

In this appendix
we briefly show the opposite case,
the observational appearance of the black hole accretion,
which is quite different from the black hole wind.

In the accretion flow
the optical depth in the fixed frame should be evaluated as
\begin{equation}
   d\tau = \gamma (1 + \beta \cos\theta) \kappa_0 \rho_0 ds,
\end{equation}
where $\beta$ is the inward velocity.

For the spherically symmetric accretion onto a black hole,
we assume that the velocity field is a free-fall one.
Hence, both the four velocity $u$ and the normalized velocity $\beta$
are expressed as
\begin{equation}
u=\beta=\sqrt{\frac{r_{\rm g}}{R}},
\end{equation}
since $y=\gamma \sqrt{g_{00}}=1$ for a free-fall case,
$g_{00}$ being $(1-r_{\rm g}/R)$.

From continuity equation, for the spherically symmetric stationary accretion,
the rest mass density $\rho_{\rm 0}$ measured in the comoving frame
varies as  
\begin{equation}
  \rho_{\rm 0} = \frac{\dot{M}}{4 \pi} \frac{1}{R^{2}u},
\end{equation}
 where 
$\dot{M}$ is the mass-accretion rate.

Furthermore, we suppose the temperature $T_0$ of 
the accretion gas in the comoving frame varies as
\begin{equation}
   T_0 = \left( \frac{\dot{E}}{4\pi \sigma R^2} \right)^{1/4},
\end{equation}
where
$\dot{E}$ is the energy-outflow rate (luminosity) in the comoving frame.

For the present problem
of the observed appearance of optically thick accretion,
such a photon trapping effect as well as the space-time curvature
are not significant,
since we may see the very outside of the accreting flow.

Using the similar procedure in the text,
we calculate the location of the apparent photosphere,
the apparent temperature, and the apparent luminosity
(figures 6-8).

\begin{figure}
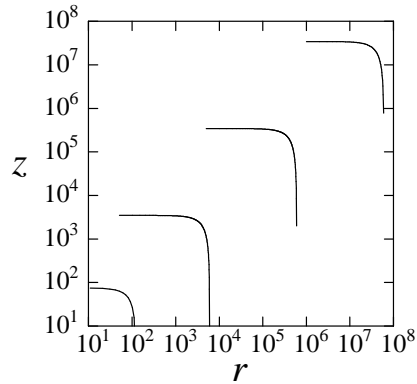

  \begin{center}
  \FigureFile(80mm,80mm){figure06.eps}
  \end{center}
\caption
{
Location of the apparent photosphere of a black-hole accretion
for various mass-accretion rates.
The mass-accretion rates $\dot{m}$ are
10, 100, 1000, and 10000 from inner to outer curves. 
The units of the $r$- and $z$-axis is the Schwarzschild radius $r_{\rm g}$.
}
\end{figure}

In contrast to the case of a black hole wind,
the apparent photosphere of a black hole accretion
generally becomes very large (figure 6).
Moreover,
it becomes larger and larger, as the mass-accretion rate is large.
This is because in the free-fall case
the infall velocity is very small at large distance
and then the density is large at large distance.
If the velocity is sufficiently small,
the height $z_{\rm ph}(0)$ of the apparent photosphere 
at $r=0$ ($\theta = 0$) is approximately expressed as
\begin{equation}
  \frac{z_{\rm ph}(0)}{r_{\rm g}} \sim \dot{m}^2.
\end{equation}
Hence, the size of the apparent photosphere of a black hole accretion
roughly expands in proportion to the square of the mass-outflow rate.
This is consistent with the present result in figure 6.

\begin{figure}
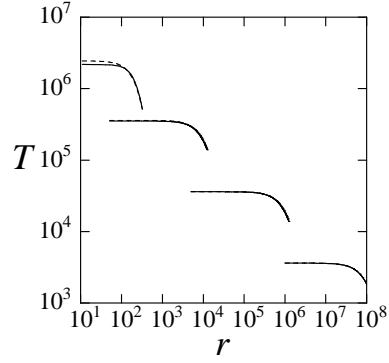

  \begin{center}
  \FigureFile(75mm,75mm){figure07.eps}
\end{center} 
\caption{
Temperature distribution at the apparent photosphere
as a function of radius $r$
for various mass-accretion rates.
The dashed curves are the temperatures in the comoving frame,
while the solid ones are those in the fixed (observer's) frame.
The mass-accretion rates $\dot{m}$ are
10, 100, 1000, and 10000 from top-left to bottom-right.
}
\end{figure}

On the contrary,
the temperature becomes lower and lower,
as the mass-accretion rate is large,
as is shown in figure 7.
This is because we set $\dot{e}=1$.
In addition,
except for $\dot{m}=10$,
the observed temperature is almost equal to the comoving temperature.

Finally,
both the observed and comoving luminosities
are on the order of the Eddington one (figure 8).
Only in the case of $\dot{m}=10$,
the observed luminosity is smaller than the comoving one
due to the relativistic de-boost.

\begin{figure}
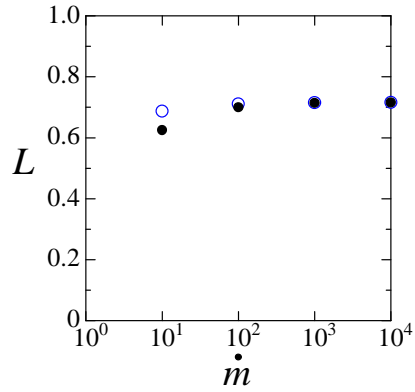

  \begin{center}
  \FigureFile(80mm,80mm){figure08.eps}
  \end{center}
\caption{
Luminosity of a black hole accretion as a function of 
mass-accretion rates $\dot{m}$.
The filled circles represent the observed luminosity in the inertial frame,
 and the open ones show the comoving luminosity.
}
\end{figure}


\end {document}